\documentclass[prd,twocolumn,showpcs,amsmath,amssymb,nofootinbib,preprintnumbers,balancelastpage]{revtex4-2}
\pdfoutput=1
\usepackage{epsfig}
\usepackage{amsmath}
\usepackage{bm}
\usepackage{times}
\usepackage{graphicx}
\usepackage{color}
\usepackage{slashed}
\usepackage{graphicx}
\usepackage{amsmath}
\usepackage{hyperref}
\usepackage[normalem]{ulem}

\def\nn{\nonumber}
\def\bea{\begin{eqnarray}}
\def\eea{\end{eqnarray}}
\def\ba{\begin{eqnarray}}
\def\ea{\end{eqnarray}}
\def\be{\begin{equation}}
\def\ee{\end{equation}}

\begin{document}

\title{The Universe Originating from an Empty Planck-Size Torus}

\author{Bartosz~Fornal}
\affiliation{Department of Chemistry and Physics, Barry University, Miami Shores, Florida 33161, USA}

\date{\today}

\begin{abstract}
We consider a Universe with a three-torus topology which before inflation is devoid of any matter or radiation. Its pre-inflationary evolution is driven solely by Casimir energies of the existing fields, with a radiation-like equation of state.  We show that, under natural assumptions and with the appropriate number of fermions and bosons in the theory, such a Universe can start its evolution from a Planck size at the Planck time. Moreover, assuming typical parameters for inflation and reheating, the three-torus Universe can be several Hubble radii across at present time, which is precisely the size  hinted by the anomaly in the cosmic microwave background observed at low multipole moments. 
We derive a relation between the size of the Universe, the number of $e$-folds of inflation, and the energy density decrease during reheating, which we then use to determine the parameter values consistent with  the anomaly and the lower bound on the size of the Universe from the Planck satellite.
\vspace{4mm}
\end{abstract}

\maketitle
\bigskip

\noindent
{\bf{1. Introduction}}
\vspace{2mm}

The question about the size and shape of the Universe is one of the most intriguing open problems  in modern  cosmology.  
Although astrophysical observations provide accurate information about  the geometry of spacetime, its topology remains a mystery.  Moreover, it is not even known if the Universe is compact, and thus finite, or infinite.

Currently, the best testing ground for the topology of the Universe is the cosmic microwave background (CMB), with data provided  by WMAP \cite{wmap} and Planck \cite{Planck:2015gmu,Planck:2018vyg}. A nontrivial topology would manifest itself as correlations between various regions in the sky. Searches for the resulting  repeated patterns, such as \emph{circles in the sky}, have been proposed and conducted  \cite{Cornish:1997ab,Levin:1997tu,Cornish:2003db,ShapiroKey:2006hm,Aslanyan:2011zp,Vaudrevange:2012da,Aslanyan:2013lsa,Bernui:2018wef,Aurich:2021ofm,COMPACT:2022gbl}, however, without a strong conclusion.

In this paper, we consider a Universe with a general three-torus topology. Interestingly, it was argued that such a toroidal universe should dominate the quantum creation of universes from nothing \cite{Zeldovich:1984vk, Coule:1999wg, Linde:2004nz}, although recently this claim has been challenged \cite{Guth:2025dal}. Here we will not be concerned with how a three-torus Universe was created, and we will just consider its evolution starting from the Planck time, with the appealing assumption that its initial size was around the Planck length.

The compact topology of the three-torus provides an extra contribution to the energy density of the Universe in the form of Casimir energies. The general formula for those Casimir energies was derived in \cite{Fornal:2011tw}, where it was shown that for a three-torus topology there exist micron-sized vacuum configurations stable with respect to the shape moduli, and unstable with respect to the volume modulus.

Those results were applied to the evolution of the three-torus Universe and used to calculate the particle production rate from the dynamical Casimir effect \cite{Fornal:2012kx}. In that reference it was shown that, starting from a Planck size and with the pre-inflationary evolution driven only by Casimir energies, assuming  72 $e$-folds of inflation ending at $10^{-33}$ s leads to the current size of the Universe of approximately ten Hubble radii.

Here we extend that analysis with a more careful treatment of reheating and by investigating the implications of the  anomaly in the CMB observed at low multipole moments, which provides hints of an upper bound on the size of the three-torus Universe.
We also leave the number of $e$-folds and the energy density drop during reheating as free parameters, using Planck's lower bounds on the size of the Universe from the search for repeated patterns in the CMB spectrum and the aforementioned upper bound suggested by the low-$\ell$ anomaly to set limits on inflation and reheating. Matching the critical energy density at Planck time leads to a striking prediction for the number of degrees of freedom at high energies.

\vspace{5mm}

\noindent
{\bf{2. Metric on the three-torus}}
\vspace{2mm}

Consider the spacetime interval of the form,
\bea
ds^2 = - d t^2 + T_{i j} d x^i d x^j\,,
\eea
where $T_{i j}$ is the metric on the three-torus $(i, j = 1, 2, 3)$ and the compact coordinates are $x^i \in [0, 1)$.
We choose the following parametrization for the metric,
\bea\label{metric_torus}
T_{i j} = \frac{b^2}{(\rho_3 \tau_2)^{2/3}}\left(
                              \begin{array}{ccc}
                                1 & \tau_1 & \rho_1 \\
                                \tau_1 & \tau_1^2+\tau_2^2 & \tau_1 \rho_1+ \tau_2 \rho_2\\
                               \rho_1 & \tau_1\rho_1+\tau_2  \rho_2& \rho_1^2+\rho_2^2+\rho_3^2 \\
                              \end{array}
                            \right) ,
\eea
where $b^3$ is the volume modulus  and  $\Phi^{\rm T}=(\tau_1, \tau_2, \rho_1, \rho_2, \rho_3)$ are the shape moduli. We assume that all those moduli depend only on time.  
This parametrization is especially convenient since $\sqrt{{\rm det}({T_{ij}})} = b^3$, which allows for a compact way of writing down the evolution equations.
The fundamental domain for a three-torus parametrized as above  is  \cite{Domain_1,Domain_2},
\bea
&&\!\!\!\!\!\!\!\!\!\!\tau_1, \rho_1 \in \left.\left(-\tfrac12, \tfrac12 \right.\right]  , \ \ \ \ 1 \leq \tau_1^2 + \tau_2^2 \leq \rho_1^2 + \rho_2^2 + \rho_3^2 \ , \nn\\
&&\!\!\!\!\!\!\!\!\!\! \ \ \ \ \ \ \ \ \ \tau_2 > 0 \ , \ \ \ \ \ \ \tau_1 \rho_1 + \tau_2 \rho_2 \leq \tfrac12 \left(\tau_1^2 + \tau_2^2\right) \ .
\eea

\vspace{3mm}

\noindent
{\bf{3. Casimir energies}}
\vspace{2mm}

Due to the compact topology, the existing fields in a three-torus Universe provide contributions to the energy density  in the form of Casimir energies.
Assuming periodic boundary conditions,  in an $N$-dimensional spacetime ($N$ was chosen to regulate the divergence) the bare contribution is given by,
\bea\label{Cas_scalar}
\rho_{0}(b, \Phi) = \sum_\alpha\frac{N_\alpha}{2\,b^3}\!\sum_{n_1, n_2, n_3 = -\infty}^\infty \hspace{-5mm}\left[(2\pi)^2{T}^{ij}n_i n_j+m_\alpha^2\right]^{\frac{N-3}{2}} \!\!\!.
\eea
In the above formula, the sum is over all particles in the theory, and $N_\alpha$ is the number of degrees of freedom for a given particle, with an extra minus sign for fermions. In the limit $N \to 4$ this formula is divergent, with the divergent part equal to minus the sum of Casimir energies of particles in flat space,
\begin{eqnarray}\label{flat}
  \rho_0^{\rm flat} =  \frac{\Gamma\left(-\frac{N}{2}\right)}{2(4\pi)^{N/2}}
  \sum_{\rm \alpha}N_\alpha \, m_\alpha^N\ .
\end{eqnarray}
This divergence is cancelled by the divergent part of the bare cosmological constant term. 

The resulting regularized formula for the Casimir energy contribution in a Universe with the topology described by  the general three-torus metric given in Eq.\,(\ref{metric_torus}) was derived in \cite{Fornal:2011tw}. 
Since we consider the evolution  during the pre-inflationary epoch, we take the limit $m_\alpha \to 0$, which yields,
\bea\label{m=0}
&&\!\!\!\!\!\!\!\!\rho_{\rm Cas}(b, \Phi) \ =\  -\Big(\sum_\alpha{N_\alpha}\Big)\frac1{b^3}\frac{{1}}{\sqrt{T^{11}}}\ \Bigg\{\frac{\pi}{6} \ T^{11}+ \frac{\zeta(3)}{2\pi}\Delta_{11}\nonumber\\
& &  + \ \ \frac{\pi^2}{90}\sqrt{\frac{{D'}^3}{\Delta_{11}}}\ \ +\,4\,\sqrt[4]{\Delta_{11} D'^{3}}\sum_{n_2, n_3=1}^\infty \left(\frac{n_3}{n_2}\right)^{\!\frac32}\nonumber\\
& & \,\ \  \times \,
 \cos\left[2\,\pi\,n_2\,n_3\frac{\Delta_{12}}{\Delta_{11}}\right] K_{\frac32}\!\left(2\,\pi \,n_2 \,n_3\,\sqrt{\frac{D'}{\Delta_{11}}}\right)\nonumber\\
& & \!+\,  2\sqrt{T^{11}}\!\!\!\!\!\!\sum_{n_2, n_3 = -\infty}^\infty\!\!\!\!\!\!\!'\ \ \ \,\sum_{n_1=1}^\infty \frac{1}{n_1}\cos\Big[\frac{2\pi n_1}{T^{11}} (n_2 T^{12}+n_3 T^{13})\Big]\nonumber\\
& & \ \ \,\times \,\sqrt{d(n_2, n_3)}\,\,\,K_1\!\left[\frac{2\pi n_1}{\sqrt{T^{11}}}\sqrt{d(n_2, n_3)}\right]  \Bigg\}\ ,
\eea
where $K_n(x)$ is the modified Bessel function of the 2\textsuperscript{nd} kind, 
\bea
\!\!\hat{\Delta} = \frac{1}{T^{11}}\Bigg(\!
  \begin{array}{cc}
    T^{11} T^{22}-(T^{12})^2 \ &\  T^{11} T^{23}-T^{12}T^{13} \\
    T^{11} T^{23}-T^{12}T^{13} \ &\  T^{11} T^{33}-(T^{13})^2 \\
  \end{array}\!
\Bigg),
\eea
\vspace{-5mm}
\bea
d(n_2, n_3) = \left(\!
                                                                                         \begin{array}{cc}
                                                                                           n_2 & n_3 \\
                                                                                         \end{array}\!
                                                                                       \right)
\hat{\Delta}\left(\!
                                                                                    \begin{array}{c}
                                                                                      n_2 \\
                                                                                      n_3 \\
                                                                                    \end{array}\!
                                                                                  \right),
\eea
and $D' = \det(\hat{\Delta})/\Delta_{11}$.  It is evident from Eqs.\,(\ref{metric_torus}) and (\ref{m=0}) that  the Casimir energy density behaves like $\sim 1/b^4$. Therefore, if only Casimir energies were present in the pre-inflationary Universe, the expansion rate was as in the radiation era, provided that the shape moduli remained constant, which we will demonstrate below.

\vspace{5mm}

\noindent
{\bf{4. Equations of motion for the moduli}}
\vspace{2mm}

Since the moduli depend only on time, upon integrating over the space coordinates the action  takes the form,
\bea\label{action}
\!\!\!\!\!\!\!S &\!\!=\!\!\!&\int \!d t \bigg[\frac{M_P^2}{2}\left(-6 \,b \, {\dot{b}^2} + b^3\dot{\Phi}^{\rm T} \hat{K} \dot{\Phi}
\right) - b^3 \rho_{\rm Cas}(b, \Phi) \bigg], 
\eea
where $M_P = 2.4 \times 10^{18} \ \rm GeV$ is the reduced Planck mass, the dot indicates derivatives with respect to time, and the  $5\times5$  matrix $\hat{K}$ has the following nonzero entries,
\bea
&&K_{11} = \tfrac{\rho_2^2+\rho_3^2}{2\tau_2^2\rho_3^2}\ , \ \ K_{22} =  \tfrac{3\rho_2^2+4\rho_3^2}{6\tau_2^2\rho_3^2}\ ,\ \ K_{55} = \tfrac{2}{3\rho_3^2}\ ,\nonumber\\
&&K_{25} = K_{52} = -\tfrac{1}{3\tau_2\rho_3}\ , \ \ K_{33} = K_{44} = \tfrac{1}{2\rho_3^2} \ , \nonumber\\
&&K_{13} = K_{31} = K_{24} = K_{42} = \tfrac{\rho_2}{2\tau_2\rho_3^2}\ .
\eea
\vspace{-5mm}

\noindent
The equations of motion are obtained by varying the action with respect to the volume and shape moduli.  This results in six second order differential equations, e.g.,
\bea
\frac{\ddot{b}}{b} \!+\!  \frac{\dot{b}^2}{2b^2}\!+ \!\frac14 \dot{\Phi}^{\rm T} \hat{K} \dot{\Phi} -\! \frac{1}{6 M_P^2b^2} \frac{\partial (b^3\rho_{\rm Cas}(b,\Phi))}{\partial b} = 0  \,. \ \ 
\eea
Those are complemented by the Friedmann equation,
\bea\label{Friedmann}
{M_P^2}\left(-6\,b \,\dot{b}^2 + b^3\dot{\Phi}^{\rm T} \hat{K} \dot{\Phi}  \right) + 2\,b^3\rho_{\rm Cas}(b, \Phi) = 0 \ .
\eea

We find that, due to Hubble friction from the growing volume modulus, the evolution of the shape moduli is damped and within a few Planck times they take terminal values dependent on the assumptions for $\Phi(t_0)$ and $\dot{\Phi}(t_0)$. For the conservative choice $\dot{\Phi}(t_0) = 0$, the final values of $\Phi$ are very close to  $\Phi(t_0)$, i.e., their assumed values at Planck time.
In order to avoid this arbitrariness, we choose a different approach.

It was shown in \cite{Fornal:2011tw} that the Casimir energy density in Eq.\,(\ref{m=0}) has one global minimum in the shape moduli space, 
\vspace{-9mm}

\bea\label{special_shape}
\Phi^T = \left(\frac{1}{2}, \frac{\sqrt{3}}{2}, \frac{1}{2}, \frac{\sqrt{3}}{6}, \frac{\sqrt{6}}{3}\right)  ,
\eea

\vspace{-2mm}

\noindent
which arises from the symmetries of the Casimir energies.
Therefore, this point corresponds to the lowest energy, and one could argue that a  Universe with these shape moduli and $\dot{\Phi}(t_0) = 0$ is most likely to be created. 
With this choice of initial conditions, the shape moduli remain constant in time, and the three-torus metric takes the form,
\bea\label{special_metric}
T_{i j} = \frac{b^2}{\sqrt[3]4}\left(
                              \begin{array}{ccc}
                                2 & 1 & 1 \\
                                1 & 2 & 1 \\
                                1 & 1 & 2 \\
                              \end{array}
                            \right) ,
\eea
leaving only the evolution of the volume modulus nontrivial. 
With  this metric, the Casimir energy density becomes,
\bea\label{zeroCas}
\lefteqn{ \rho_{\rm Cas}(b) \ =\  -\Big(\sum_\alpha{N_\alpha}\Big) \frac{1}{{b^4}}\, \Bigg\{ \ \frac{\pi\sqrt{3}}{{6\sqrt[3]4}}+{\frac{2\sqrt[3]{2}}{3\sqrt{3}}}\,\frac{\zeta(3)}{\pi}+ \frac{\pi^2\sqrt[3]{2}}{180} }\nonumber\\
& & \!+ \ \frac{8\sqrt[4]{3}}{3\sqrt[6]{2}}\sum_{n_2, n_3=1}^\infty  \left(\frac{n_3}{n_2}\right)^{\frac{3}{2}}\cos\left(\pi\,n_2\,n_3\right)
\,K_{\frac{3}{2}}\!\left(\sqrt{3}\,\pi \,n_2 \,n_3\right)\nonumber\\
& & \!+ \ {\frac{4}{\sqrt[6]2\sqrt{3}}} \sum_{n_2, n_3 = -\infty}^\infty\!\!\!\!\!\!\!'\ \ \ \,\sum_{n_1=1}^\infty  \frac{1}{n_1}\cos\left[\frac{2}{3}\pi n_1 (n_2 +n_3)\right] \nn\\
& & \!\!\times\  \sqrt{n_2^2\!-\!n_2 n_3 \!+\!n_3^2} \  \ K_1\!\left(\frac{4\pi\sqrt2}{3} \,n_1\sqrt{n_2^2\!-\!n_2 n_3 \!+\!n_3^2}\right)\!\Bigg\} \nn\\[5pt]
&& \ \  \approx  \ {0.81\,(n_F - n_B)}/{b^4} \ ,
\eea
where $n_F$ and $n_B$ are the total number of fermionic and bosonic degrees of freedom, respectively. A similar numerical factor is obtained for other choices of the three-torus metric, e.g., assuming $T_{ij} = b^2 {\rm diag}(1,1,1)$ the numerical factor is $\approx 0.8375$, which agrees with the results of \cite{Zeldovich:1984vk,Guth:2025dal}.

\vspace{5mm}

\noindent
{\bf{5. Evolution of the Universe}}
\vspace{2mm}

A Universe with a three-torus topology is flat, so there is no curvature term. Therefore, throughout its evolution the energy density is equal to the critical density, as governed by the Friedmann equation,

\vspace{-5mm}

\bea\label{Friedmann2}
H^2 = \bigg(\frac{\dot{b}}{b}\bigg)^2 = \frac{\rho}{3 M_P^2} \ . 
\eea

\vspace{-1mm}

\noindent
This ensures that the energy density will be critical also at the present time, as observations indicate \cite{Planck:2018vyg}.

We consider the pre-inflationary Universe empty of any matter or radiation, with its early evolution governed solely by Casimir energies\footnote{This scenario is not possible for a Universe with a closed topology, such as a three-sphere. The  energy density from the curvature  is $ \sim - M_P^2/R^2$ and dominates over the Casimir energy contribution \cite{Fornal:2011tw}. This results in the Friedmann equation $(\dot{R}/R)^2 = - 1/R^2 + \rho_{\rm Cas}(R) / (3M_P^2)$, which has no physical solutions for $R \ge \ell_P$. Therefore, the three-sphere Universe could not expand without the presence of matter or radiation. On the other hand, for a Universe with a compact open topology the evolution would be driven by the curvature term, yielding an expansion linear with time \cite{milne1935relativity}.}. 
The energy density in Eq.\,(\ref{Friedmann2}) is then just the Casimir energy density from Eq.\,(\ref{zeroCas}), and the expansion proceeds as in the radiation era \cite{Fornal:2012kx}.
Making the assumption, seemingly natural, that the extrapolation of the evolution equation yields $b(0)=0$, the Friedmann equation  gives,\footnote{The requirement $b(0)=0$ is equivalent to assuming that at Planck time $H(t_P) = 1/(2t_P)$, likely to be modified by quantum gravity effects.}
\bea\label{brad2}
b(t) \approx  (n_F-n_B)^{\frac14} \sqrt{\frac{t}{M_P}} \ .
\eea
Expressing this as $b(t) = b(t_P) \sqrt{t/t_P}$, where $t_P = 5.4 \times 10^{-44} \, \rm s$ is the Planck time, we obtain,
\bea\label{insize}
b(t_P) \approx \left[8 \pi (n_F - n_B)\right]^{\frac14}\ell_P \ ,
\eea
where  $\ell_P = 1.6 \times 10^{-35} \, \rm m$ is the Planck length. Therefore, the difference between fermionic and bosonic degrees of freedom determines the size of the Universe at Planck time.

For example, taking $n_F - n_B = 62$, as  in the Standard Model, sets the size of the Universe at Planck time to be,\footnote{As discussed in \cite{Linde:2017pwt}, if a Planck-size topologically nontrivial Universe does not collapse into a black hole within the Planck time, its chances to collapse later become small.}
\bea
b(t_P) \approx 2\pi \ell_P \ . 
\eea
This will be our assumption in the benchmark scenario discussed below. To obtain results for any theory beyond the Standard Model, the subsequent formulas for $b(t)$ need to be simply rescaled by $[(n_F-n_B)/62]^{1/4}$. 

Interestingly, when $n_F-n_B$ is small, such as in models with supersymmetry, the size of the Universe at Planck time is closer to the Planck length, e.g., for $n_F-n_B=1$ we get $b(t_P) \approx 2\, \ell_P$. One could even have $b(t_P) \approx \ell_P$, but this would require either $n_F-n_B\approx0.04$ (perhaps possible if some particles have masses at the Planck scale and their contribution is partially suppressed) or $b(t_0)=0$ for $t_0>0$.

\vspace{5mm}

\noindent
{\textit{Casimir energy epoch}}
\vspace{2mm}

As discussed above, assuming $n_F-n_B$ as in the Standard Model, the pre-inflationary evolution of the Universe is,
\bea\label{brad}
b(t) \approx 2\pi \ell_P \sqrt{\frac{t}{t_P}} \ .
\eea
This expansion rate holds until the onset of inflation, which happens when  the Casimir energy density falls below the energy density generated by the inflaton potential.

Since the final size of the Universe is independent  of when inflation starts, for illustrative purposes and without loss of generality we assume $t_{{\rm inf}, i} = 10^{-35} \ \rm s$. Then, at the beginning of inflation the Universe has the size:
\begin{itemize}
\item[$\star$] \ $b(t_{{\rm inf}, i})  \approx 1.4\times 10^{-30} \ \rm m$ \ .
\end{itemize}

\vspace{2mm}

\noindent
{\textit{Inflation}}
\vspace{2mm}

Since the three-torus has zero curvature, there is no flatness problem. However, inflation  \cite{Guth,Linde:1981mu,Starobinsky:1980te,Steinhardt} is still needed to solve the horizon and monopole problems (although the latter might be absent at all in the case of a three-torus Universe starting its evolution from Planck size at Planck time \cite{Uzan:1996yg}). 
Throughout its duration, the expansion is exponential, driven by the constant energy density of the inflaton field. Denoting by $N$ the number of $e$-folds, this epoch lasts for
\bea\label{tif}
\Delta t_{\rm inf} = t_{{\rm inf},f} - t_{{\rm inf},i} = {2}N \,  t_{{\rm inf},i}
\eea
and culminates with reheating \cite{Kofman:1994rk}. Assuming $N=67.7$ in our benchmark scenario, at the end of the inflationary phase but before reheating, i.e., at the time $t_{{\rm inf}, f} = (2 N+1) t_{{\rm inf}, i} $ :
\begin{itemize}
\item[$\star$] \ $b(t_{{\rm inf}, f}) =e^N  b(t_{{\rm inf}, i}) \,\approx\, 35 \ \rm cm$ \ .
\end{itemize}

\vspace{2mm}

\noindent
{\textit{Reheating}}
\vspace{2mm}

Upon the onset of reheating the energy density starts falling. Let us denote by $D$ the energy density drop during reheating, i.e., between $t_{{\rm inf},f}$ and $t_{{\rm reh},f}$, 
\bea\label{dd}
D = \frac{\rho_{\rm inf}}{\rho(t_{{\rm reh}, f}) }  \ ,
\eea
which is typically in the range $D \sim 10^4 - 10^8$.
Due to the\break dynamics of the inflaton field, which oscillates and decays into radiation, the expansion is as  in a matter-dominated era, 
\bea\label{ddd3}
b(t) = b(t_{{\rm inf},f}) \left[1+\frac34\left(\frac{t-t_{{\rm inf},f}}{t_{{\rm inf},i}}\right)\right]^{\frac23} \ .
\eea
 Equations (\ref{dd}) and (\ref{ddd3}) imply that reheating ends at $t_{{\rm reh}, f} = [2N+1+4(\sqrt{D}-1)/3] \,t_{{\rm inf},i}$. Assuming in our benchmark scenario $D = 10^5$, we arrive at:
\begin{itemize}
\item[$\star$] \ $b(t_{{\rm reh}, f}) = b(t_{{\rm inf},f}) \sqrt[3]{D}\, \approx \, 16 \ \rm m$ \ .
\end{itemize}

\vspace{4mm}

\noindent
{\textit{Post-inflationary epoch}}
\vspace{2mm}

Inflation is followed by radiation domination,  during which the size of the three-torus Universe grows according to,
\bea
b(t) = b(t_{\mathrm{reh},f})
\sqrt{
1+
\frac{t-t_{\mathrm{reh},f}}{t_{\mathrm{inf},i}\sqrt{D}}
     } \ .
\eea
This era lasts to around $t_{\rm rad} \approx 50\,000$ years after the Big Bang, and at its end:
\begin{itemize}
\item[$\star$] \  $b(t_{\rm rad}) \approx b(t_{\mathrm{reh},f})
{D}^{-\frac14}\sqrt{
{t_{\rm rad}}/{t_{\mathrm{inf},i}}}
      = 3.6\times 10^{23} \ \rm m$ .
\end{itemize}

In the subsequent matter domination era, the expansion is,
\bea
b(t) = b(t_{\mathrm{rad}})
\left[1+
\frac{3}{4}\left(
\frac{t-t_{\mathrm{rad}}}{t_{\mathrm{rad}}}\right)
\right]^{\frac23}.
\eea
It lasts to approximately $t_{\rm mat} \approx 10$ billion years after the Big Bang and results in a Universe with:
\begin{itemize}
\item[$\star$] \  $b(t_{\rm mat}) \approx 
b(t_{\mathrm{rad}})
 \left[{3\,t_{\rm mat}}/{(4\,t_{\rm rad})}\right]^{\frac23}
= 1.0\times 10^{27} \ \rm m$ .
\end{itemize}

During the currently  ongoing dark energy domination era, the energy density remains roughly constant  and 
 the size of the Universe is growing exponentially, although with a tiny exponent. In our benchmark scenario for $N=67.7$ and $D = 10^5 $,  the current size of the Universe, i.e., at ${\mathbb T} = 13.8 $ billion years after the Big Bang, is:
 \begin{itemize}
\item[$\star$] \ $b(\mathbb T) \approx  1.3 \times 10^{27} \ \rm m$ \ , 
\end{itemize} 
consistent with the lower bound  from the Planck's search for repeated patterns in the sky \cite{Planck:2015gmu}.

\vspace{4mm}

\noindent
{\textit{Full evolution}}
\vspace{2mm}

Combining all the equations, the current size of the three-torus Universe that started with dimensions on the order of the Planck length at Planck time, is given by,
\bea\label{full}
b(\mathbb T) & \approx & \left[8 \pi (n_F - n_B)\right]^{\frac14} \ell_P \,e^N D^{\frac1{12}} \sqrt{{\frac{t_{\rm rad}}{t_P}}}  \left({\frac{3\,t_{\rm mat}}{4\,t_{\rm rad}}}\right)^{\!\!\frac23} \nn\\
&\times &e^{\sqrt{\frac{\rho_\Lambda(\mathbb T)}{3M_P^2}}({\mathbb T} - t_{\rm mat})},
\eea
where $\rho_\Lambda(\mathbb T)$ is the dark energy density today.
The time of the onset of inflation $t_{{\rm inf},i}$ cancels out from Eq.\,(\ref{full}). This leaves  only two free parameters: the number of $e$-folds during inflation ($N$) and the energy density drop during reheating ($D$).
Below we derive the constraints imposed by Eq.\,(\ref{full})  on  the parameter space $(N,D)$, given Planck's lower bound on $b(\mathbb T)$ and  the hint of an anomaly at low multipoles.

\vspace{5mm}

\noindent
{\bf{6. Implications of CMB data}}
\vspace{2mm}

Measurements of the CMB such as those made by WMAP and Planck \cite{wmap, Planck:2015gmu,Planck:2018vyg} offer an exceptional opportunity to probe the size and shape of the Universe.  The search for repeated patterns provides a lower bound on its size, whereas the observed lack of correlations at large angular scales, i.e., low multipoles, hints towards a finite size and compact topology.

\begin{figure}[t]
\centerline{\scalebox{0.62}{\includegraphics{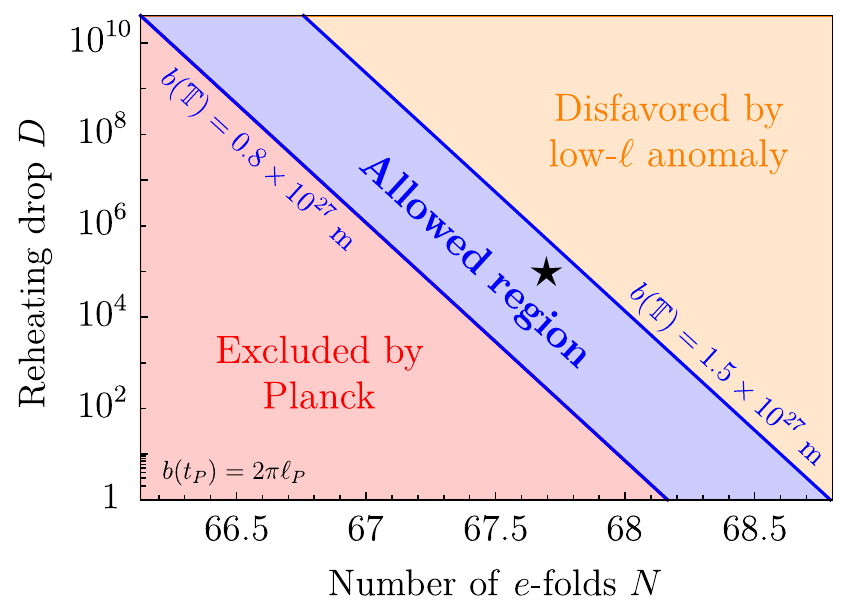}}}
\caption{\footnotesize{The allowed parameter space  (shaded in blue) for the number of $e$-folds of inflation $N$ vs.\,energy density drop during reheating $D$, 
for a three-torus Universe starting its evolution from the size $2\pi \ell_P$ at Planck time (corresponding to $n_F - n_B = 62$) and with its shape described by metric (\ref{special_metric}). The red shaded region is excluded based on searches for repeated patterns in the CMB, whereas the orange shaded region  is not favored by the low multipole anomaly present in the  data. The star indicates the benchmark point discussed in the text with $N=67.7$ and $D = 10^5$, for which the current size of the Universe is $1.3\times 10^{27}\ \rm m$.\vspace{1mm}}}\label{fig:1}
\end{figure}

Planck's search for matched circles in the sky yields the lower bound on the radius of the largest sphere inscribed in the fundamental domain as $R_i > 0.97 \,\chi_{\rm rec} \simeq 3 R_H$ \cite{Planck:2015gmu}, where $\chi_{\rm rec}$ is the distance to the recombination surface and $R_H$ is the Hubble radius. For a three-torus described by the metric in Eq.\,(\ref{special_metric}), this bound translates to
\bea\label{con1}
b(\mathbb T) \gtrsim 0.8 \times 10^{27} \ \rm m \ .
\eea

Observations of the CMB carried out by WMAP and Planck consistently show a  quadrupole moment lower than expected from the $\Lambda$CDM paradigm, which is part of a broader low-multipole suppression  seen in the data. The quadrupole $C_2$ is essentially governed by the Sachs-Wolfe effect,
\bea
C_2 \sim \sum_{n\ne 0} P_\Phi (k)\left[\, j_2(k \chi_{\rm rec})\right]^2 \ ,
\eea
where $k^2 = (2\pi)^2 T^{ij} n_i n_j$,  $P_\Phi (k)$ is the power spectrum, and $j_2(x)$ is the spherical Bessel function of the first kind, with its first maximum located at $k \chi_{\rm rec} \approx 3.3$. As a result,  the largest contribution to the quadrupole moment comes from the modes $k \chi_{\rm rec} \sim 2.0 - 4.6$. 

A 
 three-torus topology of the Universe can suppress this leading contribution to the quadrupole moment if the infrared cutoff enters this window, i.e., $k_{\rm min} \chi_{\rm rec} \sim 2.0 - 4.6$, since  
 then the very long wavelengths  can no longer be supported. 
 We find that 
for our metric (\ref{special_metric}) this suppression happens when  $1.5\,\chi_{\rm rec} \lesssim b(\mathbb T) \lesssim 3.4 \,\chi_{\rm rec}$, or, equivalently,
\bea\label{con2}
0.6 \times 10^{27} \ {\rm m}  \lesssim b(\mathbb T) \lesssim 1.5 \times 10^{27} \ \rm m \ .
\eea

Combining the bounds in Eqs.\,(\ref{con1}) and (\ref{con2}), we arrive at the following preferred size of today's Universe,
\bea\label{con3}
b(\mathbb T) \in (0.8 - 1.5) \times 10^{27} \ \rm m \ .
\eea

This range for $b(\mathbb T)$ can now be used with Eq.\,(\ref{full}) to derive constraints on $(N, D)$. Our results are shown Fig.\,\ref{fig:1}, with the allowed region shaded in blue. The region shaded in red is excluded by circles in the sky searches, whereas the region shaded in orange is disfavored by the low-$\ell$ anomaly.  The benchmark point corresponding to $N=67.7$ and $D = 10^5$, discussed earlier, is denoted by a  star.

Typical values for the energy density drop during reheating in models discussed in the literature are in the range $D \sim 10^4 - 10^8$, which narrows down the required number of inflation $e$-folds to $N \sim 66.6 - 68.0$. 

On the other hand, instantaneous reheating ($D=1$), for which the entire vacuum energy gets converted into radiation, requires the number of $e$-folds to be in the range,
\vspace{-5mm}

\bea
N  \in (68.2, 68.8) \ .
\eea
\vspace{-5mm}

In models with a low reheating temperature, the energy density drop can be enormous, corresponding to $D \gg 10^{10}$. Although such values are not shown in Fig.\,\ref{fig:1}, the corresponding bounds on the number of inflation $e$-folds can be obtained by extrapolating the blue lines of the allowed region.

Finally, we note that although the constraints in Fig.\,\ref{fig:1} are plotted for $n_F-n_B = 62$, an equivalent plot for a different case is obtained by shifting the allowed region horizontally by $\Delta N = - \frac14\ln[(n_F-n_B)/62]$. For example, if $n_F-n_B=1$, i.e., the Universe at Planck time had the size $b(t_P) \approx 2 \,\ell_P$, the  blue allowed band in Fig.\,\ref{fig:1} moves horizontally to the right only by $\Delta N \approx 1$, indicating that the change  is not substantial.

\vspace{5mm} 

\noindent
{\bf{7. Conclusions }}
\vspace{2mm}

It is quite remarkable that Casimir energies in a Universe with a three-torus topology can alone account for  the  energy density at Planck time and throughout the pre-inflationary epoch.
In such a scenario, under natural assumptions, the difference between fermionic and bosonic degrees of freedom   in the high energy extension of the Standard Model at the Planck scale determines the size of the Universe at the Planck time. For example, if $n_F - n_B =1 $, then at Planck time the size of the Universe is approximately two Planck lengths.
\vspace{1mm}

We traced the evolution of the volume modulus throughout the different epochs, expressing the current size of the Universe in terms of only two  parameters: the number of $e$-folds during inflation and the energy density drop during reheating. For a natural choice of values for those parameters, a three-torus Universe that started from a size on the order of the Planck length at the Planck time can have several Hubble radii across today. 
\vspace{1mm}

Using the results of existing  searches for repeated patterns in the sky and investigating the suppressed contribution to the quadrupole moment in the CMB spectrum, we determined the preferred range for the size of the  Universe and the resulting constraints on inflation and reheating, which are very restrictive. It would be interesting to investigate the impact of the three-torus topology with the metric considered here on other low-$\ell$ moments, especially since the numerical analyses conducted in the literature assumed only a rectangular torus. 
\vspace{1mm}

In   \cite{Vaudrevange:2012da,Aslanyan:2013lsa,Bernui:2018wef,COMPACT:2022gbl} it  was shown that the best fit to the CMB data is provided by the $\mathbb{R}^2 \times S^1$ topology, where the diameter of the circle is $1.4\,\chi_{\rm rec}$ today. In the spirit of our work, one could investigate the constraints on inflation and reheating obtained by starting the evolution from a Planck-size circle and ending it on this best fit value. The formulas for Casimir energies simplify considerably in that case.
\vspace{1mm}

Very recently it has been demonstrated in  \cite{COMPACT:2026vdj} that Casimir energies in a Universe with a three-torus topology may have an impact on the post-inflationary anisotropy, providing novel measurable signals in the CMB. Since the calculation was done for a rectangular  torus, it would be intriguing  to investigate whether the metric considered here provides additional contributions to those anisotropies.

\bibliography{bibliography}

\end{document}